# An easy technique for focus characterization and optimization of XUV and soft x-ray pulses

A. A. Muschet[1], A. De Andres[1], N. Smijesh[1,2] and L. Veisz[1,*]

[1] Department of Physics, Umeå University, SE-90187 Umeå, Sweden
[2] Ultrafast Optics Group, School of Pure and Applied Physics, Mahatma Gandhi University, 686 560, Kottayam, Kerala, India
[*] Correspondence: laszlo.veisz@umu.se

**Abstract:** For many applications of extreme ultraviolet (XUV) and x-ray pulses a small focus size is crucial to reach the required intensity or spatial resolution. In this article, we present a simple way to characterize an XUV focus with a resolution of 1.85 µm. Furthermore, this technique is applied for the measurement and optimization of the focus of an ellipsoidal mirror for photon energies ranging from 18 – 150 eV generated by high-order harmonics. We envisage a broad range of applications of this approach with sub-µm resolution from high-harmonic sources via synchrotrons to free-electron lasers.

## 1. Introduction

For a high spatial resolution on the µm scale [1,2] as well as nonlinear interactions between XUV and soft-x-ray pulses and matter [3], which are a necessary prerequisite for e.g. XUV pump – XUV probe experiments, a small and well characterized focus is of crucial importance [4]. Furthermore, a high-energy throughput of the focusing system is essential, to get the energy of the XUV source onto the experimental target. For a bandwidth of a few eV a small focus can be achieved via the use of multilayer coated spherical mirrors in normal incidence geometry. However, for a broad bandwidth those multilayer mirrors have low overall reflectivity [5–7]. Therefore, a high energy throughput is typically achieved via the use of grazing incidence focusing optics [8,9]. The ideal mirror shape to image a source onto a focus spot is given by an ellipsoidal mirror, but those mirrors are very difficult to produce with high surface quality [10,11]. Therefore, combinations of multiple mirrors that are easier to produce, like toroidal mirrors, are often used in a geometry that reduces the aberrations [4,9,12,13]. Grazing incidence optics are very difficult to align, due to their high asymmetry and the fact that small misalignments cause large aberrations [14]. However, if aligned well, those mirrors can deliver focus sizes on the single micrometer scale with a high-energy throughput throughout the entire XUV and deep into the soft-x-ray spectral region.

To enable an efficient optimization of the mirror alignment, a fast and well-resolved measurement of the XUV / soft x-ray focus is crucial. Therefore, many different methods are nowadays used to accomplish this goal.

One way to characterize an XUV / soft x-ray focus is sending it into a gas medium and measure the distribution of the generated ions. Ion based characterization methods provide a very high resolution down below one µm [15]. However, they typically measure the projection of the XUV focus along a direction normal to the propagation direction [16,17]. Due to the fact, that aberrations of XUV beams often do not fulfill circular symmetry, the information that is lost due to the projection makes the optimization of the focus difficult. Aberrations within the projection direction cannot be seen, which can easily lead to a false characterization and optimization of the focus. Furthermore, typical ion detectors need high or ultrahigh vacuum as well as high voltages, which make them expensive and cumbersome to use. In addition, the number of ions that are generated per laser pulse is limited by space charge effects. This means that the ion signal has to be averaged over many XUV pulses to achieve a good signal-to-noise ratio. This fundamentally hampers single-shot focus measurements.

Another common technique for the measurement of the size of a focus is the knife-edge technique. Here, the energy after the focus is determined while a well-defined straight beam block (referred to as knife-edge) is moved through the focus with very high precision. This method provides very high resolution, well below the µm scale [18,19] and it only needs a precise linear stage and an XUV detector, e.g. an XUV photodiode or an XUV charge-coupled device (CCD) camera. However, a measurement with the knife-edge technique gives also only information about the projection of the focus. Furthermore, a scan of the position of the knife-edge is needed to obtain this data, therefore, single-shot measurements are not possible. It cannot be used to measure the shot-to-shot fluctuations of the focus, and a large pointing fluctuation falsifies the spot size measurement.

The measurement of the focus of an XUV / x-ray beam can also be realized with the use of a scintillator [20–23]. This scintillator generates florescence wherever it is hit by the XUV photons. The photon energy of this fluorescence is typically in the visible spectral region and therefore, can be imaged with a microscope objective onto a common CCD or complementary-metal-oxide semiconductor (CMOS) camera. Typically, a resolution on the sub-10 µm scale is achieved, single-shot operation with reduced signal-to-noise ratio possible, and the fluorescence materials are rather thin (100 µm – few mm). Furthermore, a high conversion efficiency and efficient collection of the secondary photons is needed, to enable a good signal-to-noise ratio.

Wavefront sensing is another common option for the characterization and optimization of XUV and x-ray foci [12]. For this technique the profile and wavefront of the XUV beam is measured with a 2D pinhole array and a CCD outside of the focal plane and the focus is reconstructed via back-propagating algorithms. This method enables single-shot measurements, which is important for the focus optimization and the determination of the shot-to-shot stability. Furthermore, it is possible to characterize very small foci, because the measurement is performed outside of the focus, where the beam diameter is large [24]. The necessary hardware for the measurement of the XUV wavefront is commercially available, but this technique needs a very sensitive XUV camera, which is typically expensive. The reason for this is the large beam diameter at the camera position, which leads to a low fluence. Furthermore, these kind of sensors are very sensitive to internal alignment errors and often have to be calibrated via the use of a pinhole as spatial filter that blocks most of the XUV signal [24]. Moreover, this method provides an indirect characterization of the focus and, hence, is very sensitive to measurement errors. If e.g. the wavefront is not measured with high enough resolution or accuracy, this will lead to a false and typically smaller measured focus size.

Placing a sensitive XUV or x-ray CCD camera in the focus of the XUV beam would allow for a direct characterization of the XUV focus, but this is typically not a viable option, because these cameras are mounted on a vacuum flange and cannot be placed easily into the focal plane. Furthermore, their pixel size is too large (e.g. 13.5 µm or 26 µm for Andor, iKon and Newton) to enable high enough spatial resolution to characterize XUV foci on the few µm scale directly.

In this article, we describe the measurement and optimization of the XUV focus of an ellipsoidal mirror by the use of a slightly modified low-budget CMOS camera, which is commonly used for near infrared or visible light measurements. We will discuss the resolution, acquisition speed and signal-to-background capabilities at different wavelength regions from 18 – 150 eV as well as the challenges that have to be overcome to apply this technique.

## 2. Experimental implementation

For these high-harmonic generation (HHG), characterization and focus optimization experiments the attosecond beamline of the Relativistic Attosecond Physics Laboratory at Umeå University was used (see Figure 1). This beamline applies an energy upscaling approach for HHG in gas medium [25] [26]. With this approach the sub-5-fs >10 TW pulses of the laser, the Light Wave Synthesizer 20 (LWS-20), are utilized to generate intense isolated attosecond pulses with a spectrum up to 150 eV [27].

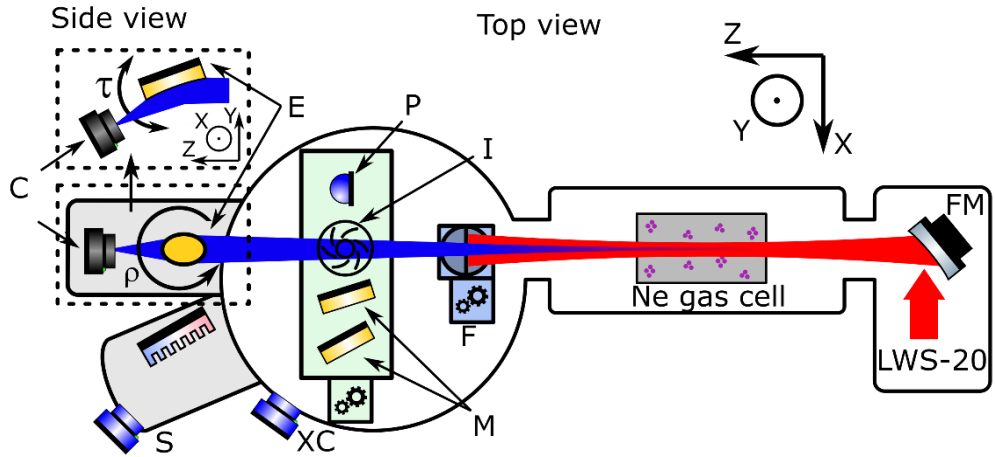

**Figure 1.** Experimental setup. LWS-20: Laser system; FM: Focusing mirror of the driving laser (focal length: 22 m); F: Filter wheel with different thin metallic foils; P: XUV photodiode to measure the XUV pulse energy; I: Motorized iris; XC: XUV CCD camera to measure the XUV beam profile; M: Gold mirrors to send the XUV beam to the XUV spectrometer and the XUV profiler; S: XUV spectrometer; E: Ellipsoidal mirror; C: CMOS camera for the measurement of the XUV focus; X, Y, Z: Defines the coordinate system as it is used in the article; τ: Grazing incidence angle of the ellipsoidal mirror; ρ: Azimuth rotation angle of the ellipsoidal mirror. There are two additional flat grazing incidence mirrors in front of the filter wheel, which are not included in this illustration.

After generation the fundamental of the laser was suppressed with thin metal foils. These metal foils act also as filters, to investigate the behavior of the focus as well as the behavior of the detection method, at three different photon energy ranges. The following filter combinations were used: 3x 150 nm Zr: ~65 – 150 eV; 2x 150 nm Zr + 1x 100 nm Pd: ~95 – 150 eV; 2x 500 nm Al: ~18 – 73 eV [28]. These energy ranges will further be referred to as Zr window, Pd window and Al window, respectively. The XUV source provides attosecond pulse energies directly at the source of 40 nJ in the Zr window, 20 nJ in the Pd window and 30 nJ in the Al window, which was determined with an XUV photodiode [16] [29]. The spectrum of the XUV source for the Zr window and the Pd window is shown in Figure 2. In addition to the spectrum and the pulse energy, the profile of the XUV pulses was measured with an XUV CCD camera (Andor, iKon), giving a full width at half maximum (FWHM) beam diameter of ~1.5 mm for all three cases.

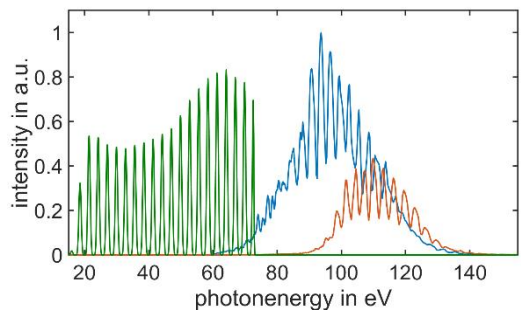

**Figure 2.** XUV spectrum of the attosecond beamline of the Relativistic Attosecond Physics Laboratory at Umeå University with 3x 150 nm Zr filters (blue), 2x 150 nm Zr + 1x 100 nm Pd filters (red) and 2x 500 nm Al filters (green). The spectrum with the Al filters was simulated using the transmission of the Al filters and a plateau odd harmonic spectrum as the spectrometer was operating in a shorter wavelength range. The area below the Al spectrum was scaled to be 50% of the area below the Zr spectrum corresponding to the energy in this region.

After spectral filtering the XUV pulses were focused with a gold coated ellipsoidal mirror, in grazing incidence geometry. The focal lengths of this mirror were 14500 mm on the incidence side and 125 mm on the exit side and the grazing incidence angle was 8°.

Considering the transmission of the filter combinations (Zr: 8%, Pd: 3%, Al: 4%) and the reflectivity of all mirrors in the XUV beamline (there were two more grazing incidence mirrors in front of the metallic filters, which are not show in Figure 1), the energy on target is 2 nJ, 1 nJ and 1 nJ for the Zr, Pd and Al window, respectively. The main cause of this low transmission was the metal filters, which have reduced transmission due to a thin oxide layer on their surfaces. The transmission of the filters was experimentally determined. New filters would increase the focused energy [28]. In experiments where high intensity is needed, e.g., ionization studies with the attosecond beamline, the use of only one metal filter is enough, which boosts the transmission by almost a factor of 10. However, for the characterization of the XUV focus, which is described in this article, the suppression of the fundamental laser radiation has to be high and therefore, the filter combinations according to the Zr window, the Pd window and the Al window are necessary.

For the detection and optimization of the XUV focus a low-budget CMOS camera with a pixel size of 1.85 µm was used (Flir, BFS-U3-120S4M-CS Blackfly S USB3, Mono). There are two windows in front of the CMOS chip to protect this type of camera from dust. For the application in the XUV spectral region those windows have to be carefully removed, to enable the XUV photons to directly hit the surface of the sensor.

Furthermore, common CCD or CMOS cameras are not designed for the use in a vacuum environment. This can presumably negatively impact the performance of the camera and increase the backing pressure inside the vacuum chamber. However, after months of operation the only negative impact of the vacuum environment on the camera that we have found is overheating. If the pressure in the vacuum chamber is <1 mbar the cooling of the camera is not sufficient anymore and it slowly heats up. Although, an increase of the noise level due to the higher temperature was not detected. After ~30 minutes of continuous operation, the camera shuts down due to too high temperature. However, when unplugged the camera cools down within <5 minutes and is useable again afterwards. No negative long-term effect for the camera have been observed from this operation. Minimal additional cooling, like the heat loss due to ~1 mbar of gas or an extra small Peltier cooling, is already enough to prevent overheating completely. As in many other XUV experiments, the focal point of our XUV beam is located in an area, where it can interact with a localized gas source. Hence, the vacuum in this part is only on the order of $10^{-5}$ mbar. At this vacuum level, the backing pressure is not negatively influenced by the implementation of the camera. For an ultrahigh vacuum environment this could be different, however, the focused XUV beam at the camera is extremely small, which gives an ideal opportunity for differential pumping, which reduces the impact of the camera onto the vacuum even further.

## 3. Results and Discussion

To minimize the effect of aberrations from the focusing mirror and achieve a small well-defined focus in the Zr window, the XUV beam diameter was reduced to ~0.8 mm in the first test. The measured focus of this beam is shown in Figure 3. This illustrates that the CMOS camera is capable of measuring signals with ~1 pixel in size and it does not show any unwanted effects like blooming under these conditions. Although, the iris clipped a significant part of the XUV beam, the signal-to-background ratio of a single-shot image was still 100 as is shown in Figure 3a-d. In this article, we use the signal-to-background ratio instead of the signal-to-noise ratio, because the root mean square (RMS) noise is well below one count. The gain of the camera was set to 1.5x amplification − full amplification scale of the CMOS camera: 1x – 16x. We estimate that 44 electrons give one count at an amplification of 1x and one photon with the central energy in the Zr window (100 eV) makes 27 electrons. The FWHM beam sizes of the x and y projections are 1.9x5.6 µm, while the equivalent circle FWHM diameter is 3.9 µm, which are resolution limited. We thereby define the equivalent circle diameter as the diameter of a circle with the same area as the sum of the area of all pixels with more or equal 50% of the maximal intensity. Furthermore, we define the energy content above half maximum as the energy in this region divided the total energy. The energy content above the half maximum with closed iris is 25%.

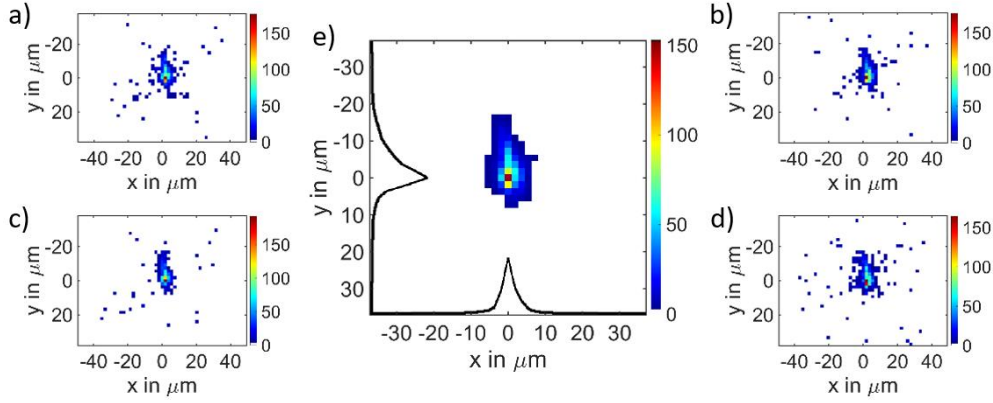

**Figure 3.** Focus of the XUV beam in the Zr window with closed iris (~0.8 mm). The resolution of the camera is only limited by its pixel size. a-d) Four single-shot images e) Average of 300 single-shot images .The black lines indicate the horizontal and vertical projections of the 2D focal image with FWHM spot sizes of 1.9x5.6 µm. The equivalent circle FWHM diameter is 3.9 µm.

With a fully open iris and the full energy of the Zr window, the gain of the CMOS camera had to be reduced to 1.1x amplification to avoid saturation with the XUV signal from a single laser pulse. Furthermore, the signal-to-background ratio increased to 150. Figure 4b shows the measured image of this XUV focus. The focus is slightly elongated in the downwards direction. We infer that this is caused by imperfections of the ellipsoidal focusing mirror, which produce aberrations in the XUV beam. Although the intensity of the signal from these aberrations is in average 10 times weaker than the main signal, it is still easily detectable. The FWHM beam sizes of the x and y projections are 7x26 µm, while the equivalent circle FWHM diameter is 3.8 µm. This result also shows that the measurement of the projections is sometimes misleading, even if both projections are observed. The energy content above the half maximum is 11%.

The high sensitivity of the CMOS camera allows the measurement of the XUV beam profile outside of the focus. Figure 4a and 4c show two such beam profiles measured 1 mm in front of and 1 mm behind the focus. Due to the high signal-to-background ratio and the resolution of 1.85 µm very fine and weak structures can be identified in those images. The aberrations introduced by the ellipsoidal mirror generate up to two side peaks in the profiles close to the focus, which are relevant with open iris. It should be mentioned that typical XUV sources based on HHG have ~100 times less pulse energy than the attosecond beamline we were using for this experiment. This decreases the signal-to-background ratio for a single-shot acquisition, especially for measurements outside of the focal plane. However, the repetition rate of those sources is typically at least 100 times higher. Hence, an image with a good signal-to-background ratio can still be acquired within less than a second.

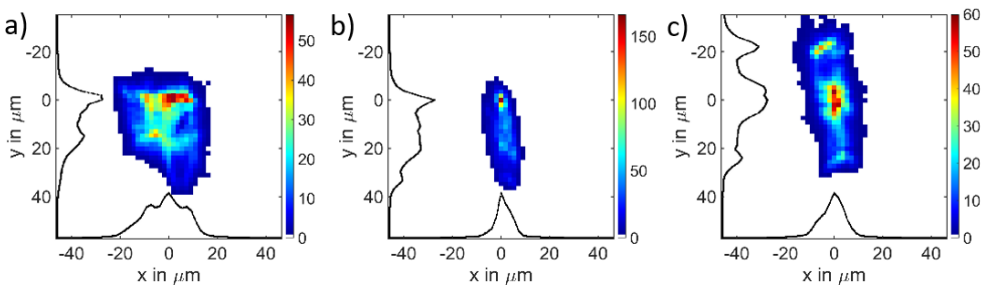

**Figure 4.** Profiles of the XUV beam in Zr window in and around the focus position with fully open iris. a) XUV beam profile 1 mm in front of the focus. b) XUV beam profile at the focal plane. c) XUV beam profile 1 mm behind the focus. 300 single-shot images were averaged for each figure. The alignment of the ellipsoidal mirror was performed in a way that the peak intensity at the focal plane was maximized. The black lines indicate the horizontal and vertical projections of the corresponding 2D image. The x and y FWHM spot sizes of the projections in the focus are 7x26 µm.

Due to this short acquisition time, the images from the CMOS camera are ideal as feedback for the alignment of the ellipsoidal mirror. Furthermore, the camera detects the 2D image of the focus without any projection. Hence, cylindrical symmetry does not need to be assumed and the errors that are caused by this assumption are fully avoided. E.g., for the focus image of Figure 4b the projection along the y axis hides the elongation of the focus in the downwards direction, whereas the projection along the x direction indicates a significantly larger focus than it actually is. All following measurements are made with fully open iris.

For the focal measurements presented in this article, the ellipsoidal mirror was aligned in a way that the peak intensity in the images was maximized for the focal plane. The short wavelength and the short focal length make the alignment of the ellipsoidal mirror very sensitive. This is illustrated in Figure 5. Figure 5a-b show the XUV beam profile at the focal plane with a misalignment of ±0.04° of the azimuth angle around the vertical axis (see Figure 2). As illustrated, this small misalignment leads approximately to a factor of 2 reduction in peak intensity. The alignment of the grazing incidence angle is even more sensitive. A misalignment of ±0.015° already leads to a decrease of more than a factor of 2 in peak intensity at the focal plane. However, it has to be mentioned that the plane with the highest intensity changes for different alignments of the ellipsoidal mirror. This is especially valid for the grazing incidence angle. This entanglement of various alignment parameters means that multiple dimensions have to be scanned repeatedly during the optimization procedure [14]. Therefore, the fast characterization of the XUV focus with the CMOS camera is essential for the alignment.

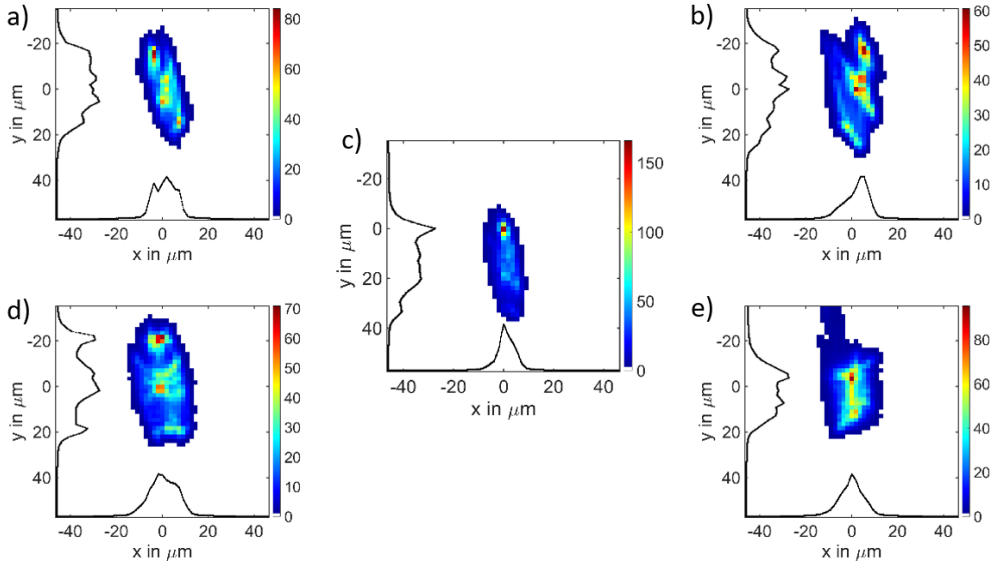

**Figure 5.** Profiles of the XUV beam at the focus position with different alignment of the ellipsoidal mirror. The rotation of the azimuth angle around the vertical axis was misaligned by a) 0.04°, b) -0.04°. c) The ellipsoidal mirror was aligned to achieve maximum peak intensity. The grazing incidence angle was misaligned by d) 0.015°, e) -0.015° with respect to the optimum. 300 single-shot images were averaged for each figure.

To evaluate the performance of the CMOS camera for different spectral regions within the XUV and soft x-ray, the focus of the ellipsoidal mirror was also observed and optimized for the Pd window and the Al window. It is known that the wavefront of high harmonics can slightly change between the harmonic orders [24]. This was also observed in this experiment, as the optimal alignment of the ellipsoidal mirror was slightly different for Zr and Al windows.

The optimized focus for the Pd window is shown in Figure 6a. The shape of this focus is very similar to the one of the Zr window. Furthermore, its peak intensity as well as its integrated signal is approximately a factor of 2 lower. This fits well with the energy measurements that were carried out with the XUV photodiode for the spectral region of the Pd window. This suggests that the spectral sensitivity of the camera does not change significantly between the spectral regions of the Zr window and the Pd window.

Figure 6b shows the optimized focus for the Al window. This focus is with an equivalent circle diameter of 10.3 µm significantly bigger than the focus of the Zr and Pd window and does not show as sharp features. This can be explained by the higher diffraction limit of the (x3) lower central photon energy that is transmitted through the Al filters. The enlargement of the focus also increases the energy content above the half maximum to 25%. Furthermore, the intensity of the camera signal was much lower and therefore, the images for the focus of the Al window had to be taken with the maximum gain of the camera (16x amplification of the signal). The signal-to-background ratio under these conditions for a single shot was 20.

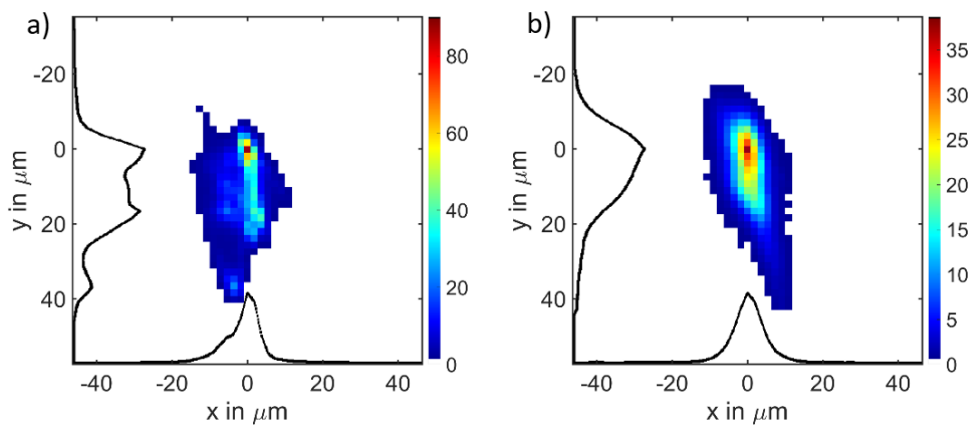

**Figure 6.** Measurement of the XUV focus at different photon energy ranges. a) Pd window (~95 – 150 eV), b) Al window (~18 – 73 eV). The x and y FWHM spot sizes of the projections in the focus for the Pd window are 6x26 µm and the equivalent circle diameter is 5.7 µm. For the Al window the x and y FWHM spot sizes of the projections in the focus are 7x22 µm and the equivalent circle diameter is 10.3 µm. 600 single-shot images were averaged for each figure.

To characterize the sensitivity of the CMOS camera in the Al window with respect to the Pd and Zr window, the XUV energy on the camera with three Zr filters was compared to the XUV energy with two Al filters. The energy with the Al filters was thereby found to be 50 ± 10% of the one with the Zr filters. After correction for the different amplification and energy, it can be concluded that the sensitivity of the CMOS camera is approximately ten times worse for the Al window than for the Zr and Pd window. This is probably due to the structure of the CMOS camera chip.

## 4. Conclusions

We describe how a common CMOS camera can be easily modified to observe XUV and soft x-ray beams. It is demonstrated that the sensitivity of this camera is high enough to acquire single-shot images of an XUV focus. This is essential for the alignment of high energy throughput, grazing incidence focusing optics, which are used in synchrotrons, free electron lasers as well as high-harmonic generation beamlines. Furthermore, the sensitivity of the camera is determined for three different spectral regions. Thereby, the sensitivity of the CMOS camera in the spectral ranges from 95 – 150 eV and 65 – 150 eV is similar, whereas the sensitivity for the spectral range from 18 – 73 eV is approximately 10 times reduced. Further improvement in spatial resolution is expected utilizing the next generation cameras with sub-µm pixel size [30].

**Funding:** L.V. acknowledges the support of the Swedish Research Council, Vetenskapsrådet (2019-02376, 2020-05111), Knut och Alice Wallenberg Stiftelse (2019.0140) and Kempestiftelserna (SMK21-0017).

**Data Availability Statement:** Data available from the authors on request.

**Acknowledgments:** We thank Roushdey Salh for the technical support.

**Conflicts of Interest:** The authors declare no conflict of interest.